\documentclass[preprint]{iucr}              

\usepackage{xcolor}
\usepackage{xspace}
\RequirePackage{graphicx}
\usepackage{epstopdf}
\usepackage{mathtools, amsmath, amssymb, amsfonts}
\usepackage{chemformula}

\usepackage{algorithm}
\usepackage{algpseudocode}
\usepackage{amsmath}
\usepackage{graphics}
\usepackage{epsfig}

\newcommand{\nmfa}{nmfMapping@PDFitc\xspace }
\newcommand{\nmfm}{nmfMapping\xspace}
\newcommand{\insitu}{\textit{in situ} }

\newcommand{\fig}[1]{Fig.~\ref{fig:#1}}

\DeclareUnicodeCharacter{2212}{\textendash}

\graphicspath{{./figures/}}

     \journalcode{A}              

\begin{document}                  



\title{{nmfMapping}: a cloud-based web application for non-negative matrix factorization of powder diffraction and pair distribution function datasets}

\author[a]{Zachary}{Thatcher}
\author[a]{Chia-Hao}{Liu}
\author[a]{Long}{Yang}
\author[b]{Brennan C.}{McBride}
\author[b]{Gia}{Thinh Tran}
\author[b]{Allison}{Wustrow}
\author[c]{Martin~A.}{Karlsen}
\author[b]{James~R.}{Neilson}
\author[c]{Dorthe~B.}{Ravnsbæk}
\cauthor[a, d]{Simon~J.~L.}{Billinge}{sb2896@columbia.edu}

\aff[a]{Department of Applied Physics and Applied Mathematics, Columbia University, \city{New York}, NY 10027 \country{United States}}
\aff[b]{Department of Chemistry, Colorodo State University, Fort Collins, Colorado 80523-1872 \country{United States}}
\aff[c]{Department of Physics, Chemistry and Pharmacy, University of Southern Denmark, DK-5230 Odense M, \country{Denmark}}
\aff[d]{Condensed Matter Physics and Materials Science Department, Brookhaven National Laboratory, Upton, NY 11973,\country{United States}}

\shortauthor{Thatcher \textbf{et al.}}







\maketitle                        

\begin{synopsis}
A web application for non-negative matrix factorization of powder XRD and PDF datasets, with examples.
\end{synopsis}

\begin{abstract}
A cloud-hosted web-based software application, nmfMapping, for carrying out a non-negative matrix factorization of a set of powder diffraction or atomic pair distribution function datasets is described.  This app allows structure scientists to rapidly find trends in sets of related data such as from \textit{in situ} and \textit{operando} diffraction experiments. The application is easy to use and does not require any programming expertise. It is available at the pdfitc.org website.
\end{abstract}


\section{Introduction}

    Modern hardware and software at x-ray synchrotrons allow scientists to study materials in ground-breaking detail \cite{maier;acie07, chupa;jac07} allowing high throughput combinatorial, \textit{in situ}, and \textit{operando} experiments that produce unprecedented amounts of data on previously inaccessible systems.

    To extract information the data need to be reduced to a scientifically meaningful function, such as a 1D powder diffraction pattern and then interpreted, for example using structural fitting such as Rietveld analysis \cite{rietv;jac69} or atomic pair distribution function (PDF) analysis \cite{egami;b;utbp12,farro;jpcm07,tucke;jpcm07}. This can be time-consuming and human intensive.  It often involves fitting the data from a single diffraction pattern at a time, as opposed to fitting cross-pattern trends, so called parametric refinement \cite{stint;jac07}.  It can often be difficult to extract trends in the data if they are not, \textit{a priori}, expected leading to missed insights.  For example,  the novel nature of modern experiments can often result in one or more unexpected structures.  Whilst tools are emerging that can automatically identify constituent structural signals \cite{yang;aca20}, they are limited to cases where the structure is in the databases they are using.  In cases when novel materials are being investigated, it is highly valuable to use structure agnostic ways of interrogating large sets of data to extract meaningful signals.

    Non-negative matrix factorization (NMF) \cite{paate;e94} has been a particularly useful, reliable, and intuitive unsupervised machine learning (ML) approach for analytically reducing large datasets of physical signals to reveal trends \cite{brune;pnas04,pauca;laa06}.  More specifically, it can yield isolated structural signals and their relative presence for both PDF \cite{liu;jac21,gedde;cc19} and powder diffraction \cite{long;rsi09} data.

    Although NMF modules are present in, for example, scikit learn \cite{pedre;jmlr11}, there are impediments to their widespread adoption.  They are not easy to use for non-programming/non-mathematician experts.  Also, even when a user is familiar with python scripting, it may not be possible or straightforward to install such libraries at any given synchrotron facility during a beamtime when such a rapid analysis of trends in data is particularly valuable.  To address these issues we have developed a cloud-hosted, web-based, application, described herein, for doing NMF decomposition of powder diffraction and PDF datasets.  This application, called nmfMapping, has been implemented as a service on the PDFitc cloud platform at pdfitc.org \cite{yang;aca21}.

\section{NMF}

    NMF has been described in depth previously in both a mathematically general sense \cite{lee;n99, berry;csda07}, and in regard to diffraction experiments \cite{liu;jac21,long;rsi09,gedde;cc19}.  We reproduce a brief discussion on the use of NMF on diffraction data here for the sake of context.

    As a matrix decomposition algorithm, NMF is a method to find and track a reduced set of basis components that can approximate the full matrix to a desired precision.  In this respect, NMF is similar to principal component analysis (PCA)\cite{lee;n99}, but with constraints applied to the decomposition that the components and the weights must be positive.  Since many experimental signals (such as intensity) are positive, and the compound signals are sums of such signals (e.g., the diffraction signals from each chemical phase in a  multi-phase sample), the enforced positivity constraints mean that NMF decomposition is more likely than PCA to find solutions that are physically meaningful. A drawback to NMF is that the optimization is non-convex and finding the global minimum is not assured.  However, in practice proper initialization seems to make NMF results consistent for diffraction and PDF data~\cite{liu;jac21,long;rsi09} where its application looks very promising \cite{todd;ic20}.

    To use it to find trends in sets of experimentally measured 1D signals we first create a matrix where each column is a signal  and each signal is measured on an identical grid of points,
    e.g., $2\theta$ for powder data or $r$ for PDF data, that form the rows.  The decomposition can then be done using standard methods \cite{pedre;jmlr11}.  Our \nmfa implementation uses standard methods but in a user-friendly web environment, with minimal preprocessing
    suited to PDF and powder diffraction data.

\section{PDFitc Interface}

    PDFitc.org \cite{yang;aca21} is a cloud based platform containing tools that can, given a user-uploaded measured PDF, automatically find candidate structures for subsequent fitting (structureMining app) \cite{yang;aca20}, predict the space group of the structure that produced the PDF (spacegroupMining app) \cite{liu;aca19}, and find the similarity between PDFs in a set (similarityMapping app).  It has an easy to navigate interface that can be utilized on all of the popular web browsers (e.g. Chrome, Firefox, Safari, etc.).  Users can log in to the web-site using Google, GitHub and ORCID credentials to upload data for processing. The website can even be utilized on a computer without data, since logging in connects users to their previous results as well as results other users may have explicitly shared with them (by default all data can only be seen by the user who uploaded it).  The goal of the website is to provide PDF analysis tools without the need to install and run software on the user's local computer.  This paper describes the addition of a new app on the PDFitc platform for carrying out NMF decomposition of a set of PDF or powder diffraction spectra collected as a related set, such as time, temperature or spatial resolved series of spectra.

\section{The nmfMapping App Interface}

    Here we describe the new NMF decomposition functionality in the nmfMapping app. We illustrate this by walking through the process of accessing the website and performing an NMF decomposition on a synthetic (calculated) PDF dataset from \cite{liu;jac21}, and then rerunning the analysis on a smart phone.	

\begin{figure}
\label{fig:pdfitc_home}
\includegraphics[width=0.8\columnwidth]{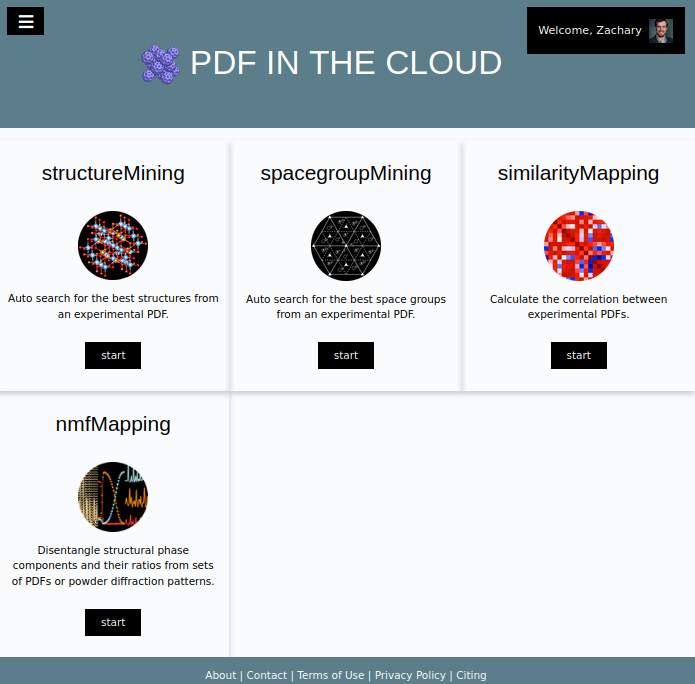}
\caption{PDFitc.org home-page.  Each tile may be clicked to access the specified app.  If not logged in, the user will be directed to a login page before continuing to the app.
}
\end{figure}

\begin{figure}
\begin{center}
    \includegraphics[width=0.8\columnwidth]{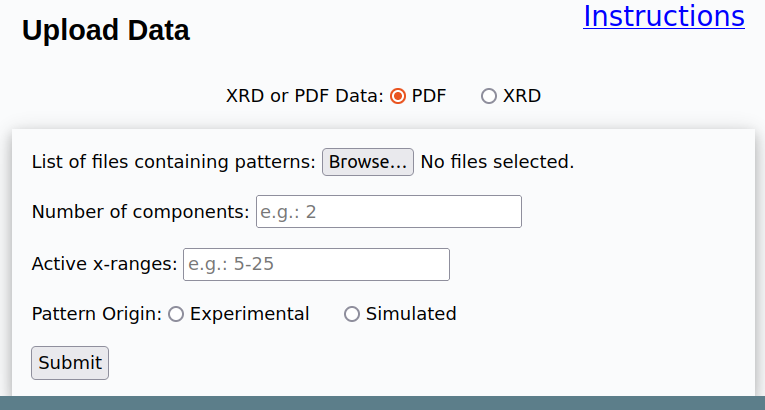}
    \label{fig:nmf_interface}
    \caption{\nmfa landing page.  Users upload their datasets via the browse button and can optionally specify parameters such as number of
    NMF components or data ranges.  After specifying if the uploaded data is experimental or simulated, the computation is started by clicking submit.}
\end{center}
\end{figure}

    On going to https://pdfitc.org the user will see the home-page of the PDFitc website (\fig{pdfitc_home}).
    Currently, including nmfMapping, PDFitc presents four application tiles (structureMining, SpacegroupMining, similarityMapping, nmfMapping) to the user.  Upon selecting the nmfMapping tile, the user will be redirected to the log-in page. Logging in is required to use the services.  It also allows the user to revisit previous analyses, as well as any analyses that have been explicitly shared by another user, through their account icon on the top-right corner of the screen, as will be shown below.  Note, a different user profile is associated with logins from each authentication service (ORCID, GitHub or Google) so it is important to pick one service and use it.  After logging in the user is directed to the nmfMapping landing page (\fig{nmf_interface}).

     To use the app a user uploads a list of either powder diffraction or PDF data, inputs a small number of configuration parameters, and hits ``submit".  The app will do the decomposition and returns the results, as described below.

     The app can currently accept \texttt{.gr} format files that are output by standard PDF analysis programs such as PDFgetX2 \cite{qiu;jac04i} and PDFgetX3 \cite{juhas;jac13}, \texttt{.chi} files that are output by integration programs such as pyFAI \cite{ashio;jac15} or Fit2D \cite{hamme;jac16}, and in general, multi-column text files such as xy or xye format files. These files may also contain header information above the columns of data.

     In terms of the configuration parameters, the most important is the ``Number of components".   Setting this to the wrong number leads to poor or non-physical results.  This value may be left blank at first and \nmfm will attempt to select the correct number of signals automatically.  The approach used is to consider the amount by which the Frobenius (frob) norm is improved by the addition of an extra contribution, divided by that index's frob norm. The number of components is increased sequentially from two until this improvement is less than 0.1\%.  If this threshold is not met, as can happen for real data with noise, an upper limit of ten components is arbitrarily imposed to get a first look at the reconstruction error graph which is returned to the user for manual consideration (elaborated on below).  The decomposition can then be rerun with the user-selected number of components after making the appropriate input to the ``Number of components" box.

     The number of components is an important parameter to control for good scientific results.  The best number roughly corresponds to the number of PDFs/powder patterns it takes to explain all the signals in the data.  For example, a sputtered film that contains overlapping patches of three metals would likely need 3 components, the three chemical components.  On the other hand a balanced chemical reaction where chemicals A and B transform to products C and D will require 2 components, even though there are four chemical components, the ``reagent signal" and the ``product signal".  If the composition of the sample does not respect the balanced synthesis equation, for example, if not all of the A chemical is consumed in the reaction, the NMF will likely require 3 components, the PDF/powder signals from the reactants, products and A.  Tuning the threshold on real data is discussed further in the examples section.  We have found that there is often an ``elbow" in the reconstruction error vs. component number graph, with it showing a distinct flattening, and this is the best first guess at the right number of components.

\begin{figure}
    \begin{center}
    \includegraphics[width=0.8\columnwidth]{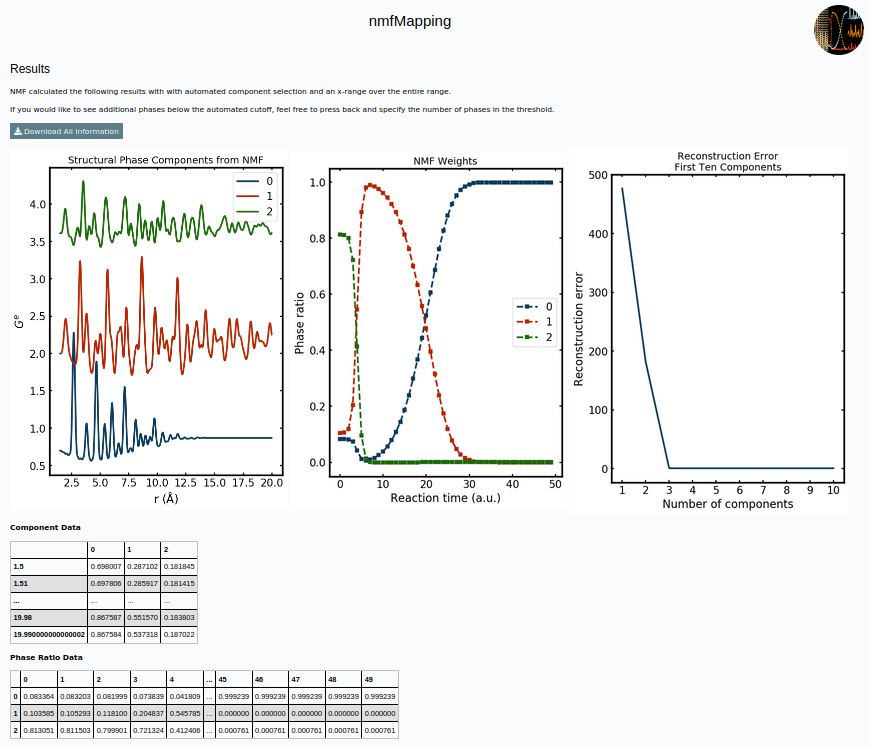}
    \label{fig:nmf_results_on_pdfitc}
    \caption{NMF Results Page after completing a decomposition.  In this case it is from the simulated PDF data described in \cite{liu;jac21}.  The first panel shows the NMF components (in this case, 3).  The center panel shows the weights that were used to recreate each uploaded PDF in the set from these three components.  The right panel is the reconstruction error vs. number of components and is useful for making decisions on how many components to try.  For this simulated and noiseless data all the error is eliminated after three components.  The tables provide a summary of the data in the figures.  All this information may be downloaded to the user's local computer.
    }
    \end{center}
\end{figure}

    After completing the decomposition, \nmfa returns three graphs and related tables (\fig{nmf_results_on_pdfitc}).
    The information in the panels is described in the figure caption of \fig{nmf_results_on_pdfitc}. They show plots of the NMF extracted components and the weights needed to reconstruct each of the the uploaded datasets.  The data in the dataset are ordered alpha-numerically by filename and so if the order is scientifically meaningful the files need to be named appropriately. In the third panel the ``Reconstruction Error" may be used to determine the right number of components to use in the NMF decomposition.  For example, in the current case, which is simulated data described in \cite{liu;jac21}, the graph shows a steep reduction in error reduction up to three components, where three is the best number of components in this case.  A number of data points from the graph are also provided in tables on the web page beneath the graphs, allowing close inspection of the data when iterating.

	All of the graphs and tables can be downloaded by the user in a compressed file, along with files containing the individual constituent components.  The output component files are in a format where they can be further processed by directly loading to most XRD or PDF fitting software, including the structureMining \cite{yang;aca20} package also located on PDFitc and the widely used PDFgui \cite{farro;jpcm07} structure refinement software.

\begin{figure}
    \begin{center}
    \includegraphics[width=0.4\columnwidth]{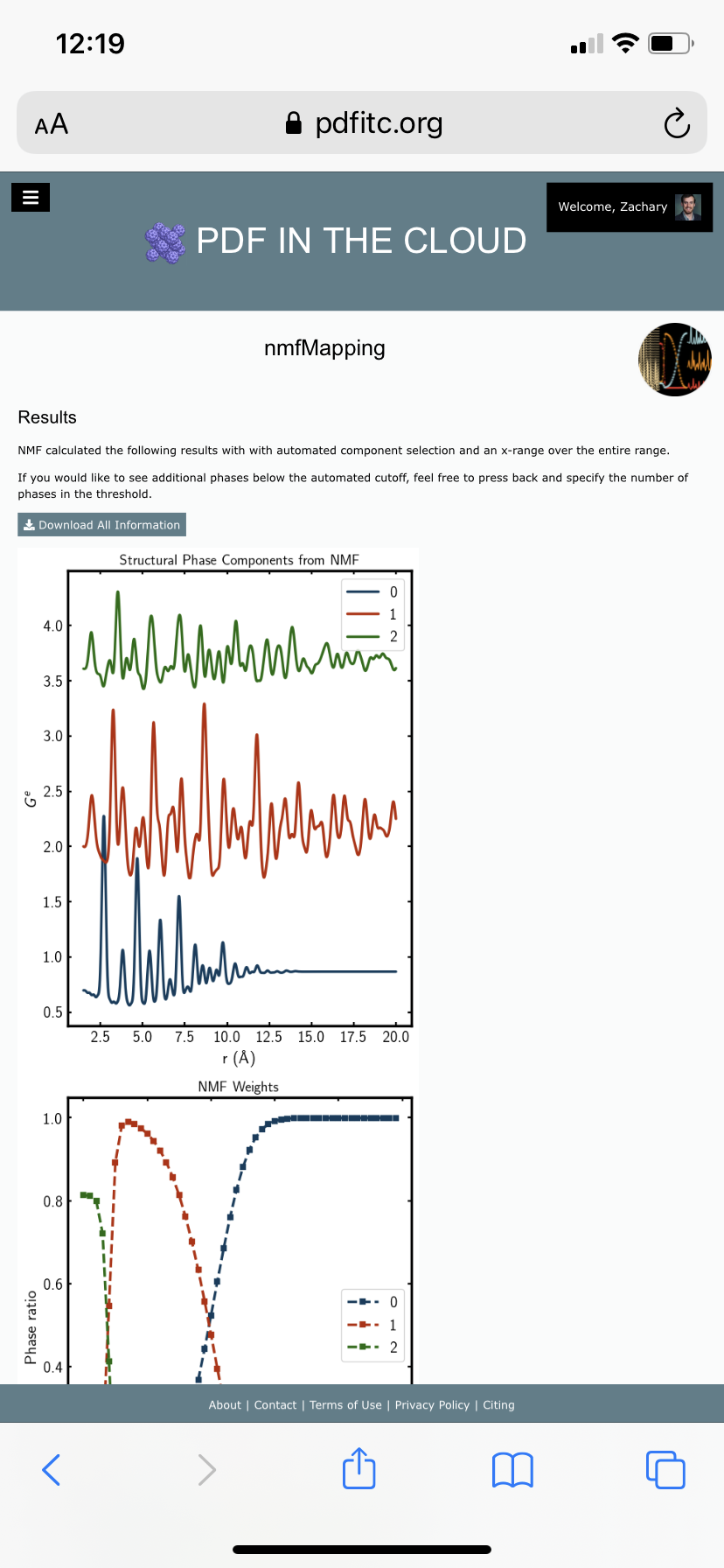}
    \label{fig:phone_screenshot}
    \caption{NMF Results Page after rerunning the analysis from the safari browser on a smart-phone.
    }
    \end{center}
\end{figure}

    \fig{phone_screenshot} emphasizes the accessibility of the \nmfa app.
After logging in and performing the analysis above on a PC where the data was located, a smart-phone was utilized to log in to the same account.  By going to the ``Activities" web page on the phone and clicking the play button, the analysis results were recreated without requiring that the data be on the smart-phone.

\section{Examples}

    In this section we present some real-world examples of \insitu experimental data run through the PDFitc app.  In the first example we consider structural changes in an \textit{operando} battery experiment, and in the second  we discuss analysis of a set of diffraction patterns collected tracking a chemical synthesis reaction.

\subsection{\emph{Operando} PDF Experiment for Rechargeable Li-ion Battery }

    In this example we consider data from an experiment characterizing nanostructured and disordered cathode battery structures via PDF analysis of \textit{operando} experiments.  Such experiments benefit greatly from use of an NMF type analysis because of the high complexity with a large number of unknown chemical components of the battery electrodes under dynamic conditions and the large numbers of datasets \cite{chris;jpe21}.

    In this particular example, \nmfa was used to quickly gain preliminary insight into the structural shifts present in a nanocrystalline \ch{TiO2}-Bronze cathode prior to a more detailed PDF fitting analysis. The x-ray data presented here were collected in transmission on a PerkinElmer (XRD1621) area detector with a wavelength of 0.20736~\AA\ at beamline P02.1, PETRA~III, DESY, Hamburg, Germany.  The battery was assembled in an AMPIX electrochemical cell configuration \cite{borki;jac12}. The battery stack consisted of a metallic lithium anode, a glass fiber separator wetted with electrolyte, and a cathode pellet made from active material (\ch{TiO2}), poly(vinylidene fluoride) (PVDF) binder, and conductive carbon.  The diffraction data was processed with DAWN~2 \cite{filik;jac17}, and PDFgetX3~\cite{juhas;jac13} to get the PDFs. More information regarding the capture and analysis of the battery cathode data, and a scientific discussion of the results, will be the subject of a future publication.

\begin{figure}
    \includegraphics[width=0.6\columnwidth]{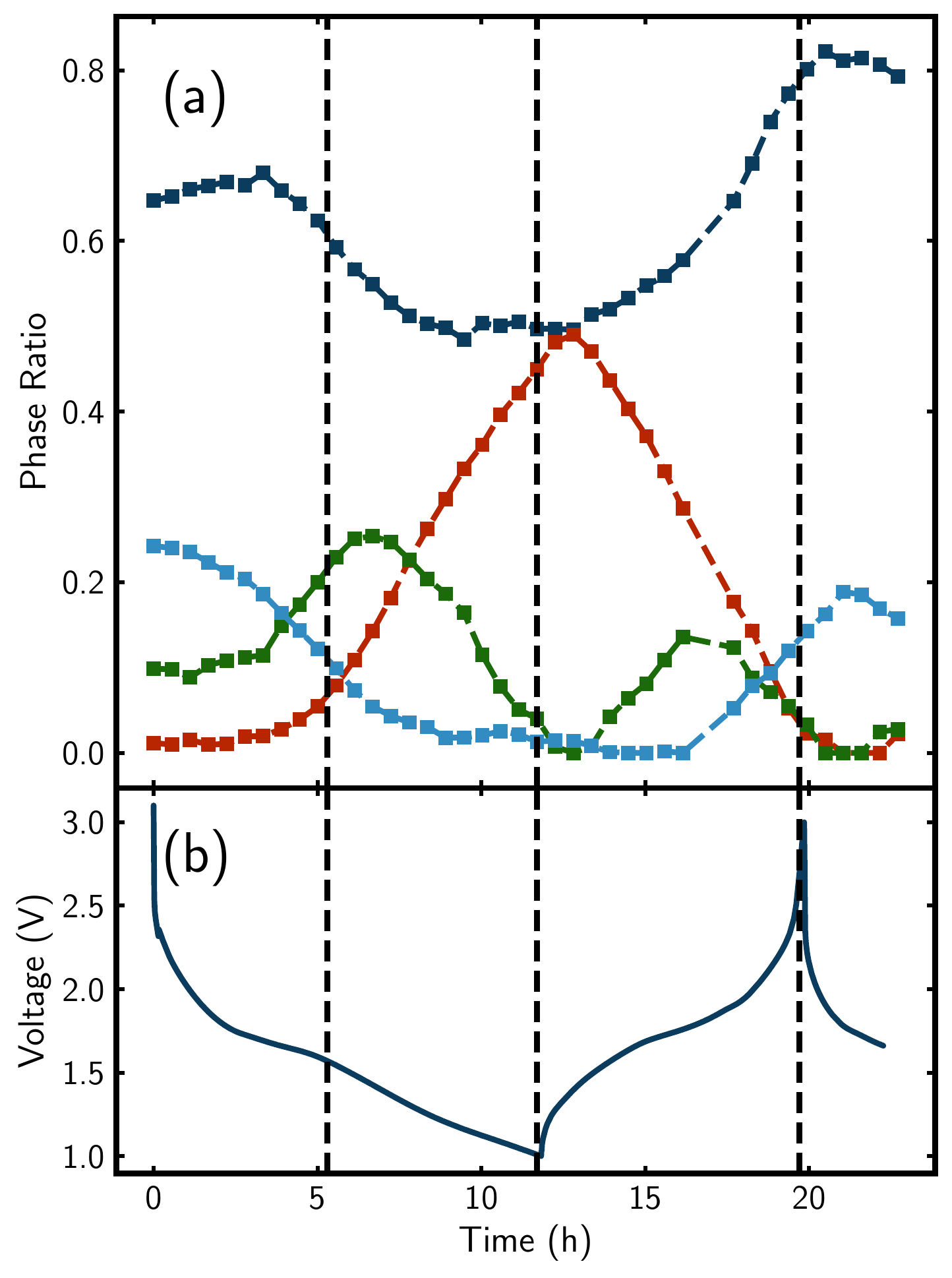}
    \label{fig:nmf_match_charge_discharge_regimes}
    \caption{(a) weights of the NMF components vs. time obtained from \nmfa and (b) the galvanostatic charge-discharge curve at the same time-points. Vertical dashed lines are drawn at significant time-points, i.e., first a change in slope of the galvanostatic curve during discharge, then the point where the cell was switched from discharge to charge, and third, the point where it was switched back again to a discharge cycle.  The NMF analysis reveals that an intermediate phase (shown in green) begins to transform into a Li-rich phase represented by the red component precisely where the voltage starts changing slope during discharge.
    }
\end{figure}

    The progress of the battery discharge is tracked using galvanostatic voltage measurements (\fig{nmf_match_charge_discharge_regimes}(b)).
The voltage over time graph reveals two regimes of different slope for the initial discharge.  The goal of the PDF measurements is to uncover any structural origin of this galvanostatic behavior.  The NMF phase tracking is also shown in \fig{nmf_match_charge_discharge_regimes}(a) lined up with the galvanostatic response for comparison.  The NMF clearly reveals that this voltage behavior is at least partially due to structural variations.  These variations are difficult to discern from the PDF, but found automatically by the NMF analysis.

\begin{figure}
    \includegraphics[width=1\columnwidth]{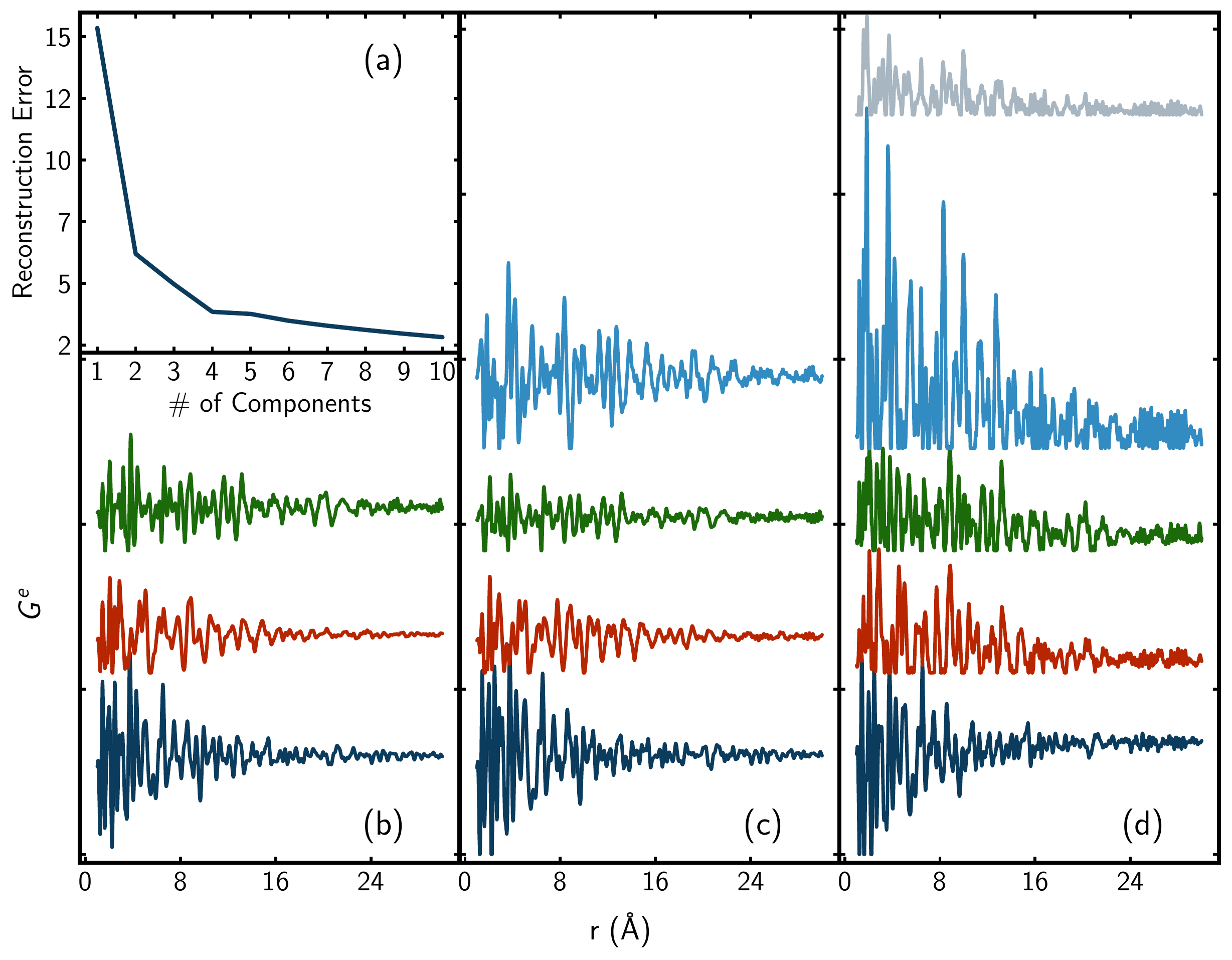}
    \caption{(a) reconstruction error vs. number of NMF components. (b)-(d) The resulting NMF components when the number of components specified in the decomposition is (b) 3, (c) 4, and (d) 5 in \nmfa .
    }\label{fig:nmf_component_noise}
\end{figure}

In order to quickly check the physical reasonableness of the \nmfa output, the analysis can be rerun repeatedly in \nmfa with different numbers of components specified.  The results of repeated NMF runs (that took only a minute or so) show that attempting to track five or more signals results in NMF components that do not resemble measured PDFs, in this case with signals that clip at high and low values (\fig{nmf_component_noise}(d)).
In this case four components appears to give the best results ((\fig{nmf_component_noise}(c)) on the basis that it is the largest number of components that appear to be undistorted PDF-like signals.  In addition,  \cite{liu;jac21} and \cite{long;rsi09} have shown that limiting the number of components to where the reconstruction error begins to flatten is a good heuristic for deciding where additional components lead to non-physical results, and in \fig{nmf_component_noise}(a) we see that this occurs with four components.   Satisfyingly, this result is also supported by chemical intuition. Two components are expected for the charged (Li-poor) state (dark blue and light blue).  At least one component (red) is expected for the discharged (Li-rich) state.  What pops out unexpectedly from this analysis is an intermediate, partially lithiated, phase (green). In general, it is important to assess the chemical validity of any NMF results, but this illustrates the utility of this approach.

\subsection{Inorganic chemical synthesis reaction}
    In this experiment a \ch{LiMnO2} and \ch{LaOCl} powder mixture was reacted at a range of temperatures in a gradient furnace \cite{onola;jac21} (see below for additional detail) for 40~min, then cooled to room temperature.   Diffraction patterns corresponding to different reaction temperatures were then measured in transmission mode at the 28-ID-2 beamline at NSLS II.  The data was integrated using GSAS-II \cite{toby;jac13} before being passed to the \nmfa app.  For more details on the experimental setup, see \cite{wustr;cm21}.

    The dataset contains 25~diffraction patterns that come from a continuum of temperatures from 300~$^\circ$C up to 750~$^\circ$C.  The gradient furnace \cite{onola;jac21} produces a more or less linear dependence of temperature vs. position along the rod-shaped sample.   The ability to measure reaction products from different temperature reactions as a function of position along the rod allows rapid acquisition of many temperatures as each data-point takes just seconds to collect.  Though in this test case there are just a manageable 25~XRD patterns it is an example of an experimental configuration with the potential to produce 100's to 1000's of patterns for a single experiment. This is a good illustration of how experimental innovation can drive the production of larger data-sets requiring more sophisticated processing such as NMF analysis.

\begin{figure}
    \includegraphics[width=1\columnwidth]{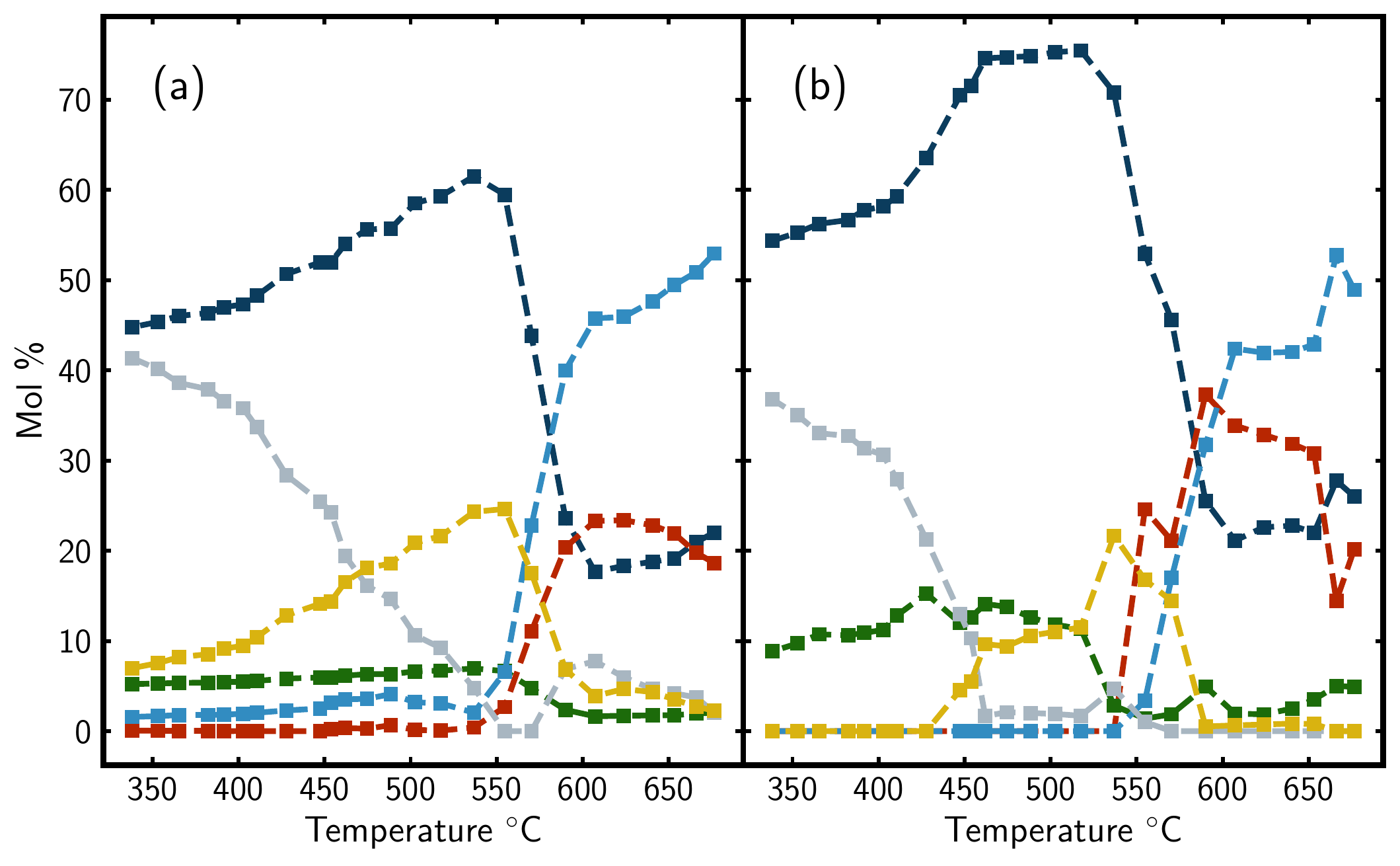}
    \label{fig:nmf_reaction_pathway_tracking}
    \caption{ Comparison of Mol \% Calculation when performed on (a) NMF component ratios and (b) sequential Rietveld fits on every dataset.  The color of the line in each graph corresponds to a specific chemical compound with the main species at low temperature being \ch{LaOCl} (dark blue) + tetragonal \ch{LiMnO2} (grey) + \ch{Mn3O4} (green). At middle temps orthorhombic \ch{LiMnO2} (gold) is formed, and at high temp the main products are \ch{LaMnO3} (light blue) + \ch{LiCl} (red), with a significant amount of left over LaOCl.
    }
\end{figure}

    One of the main goals of these experiments is to get the temperature dependence of the compositional products, which provides insight into different reaction pathways available to a system \cite{marti;cm17}.  Traditionally this would have been done by carrying out a sequence of Rietveld refinements \cite{rietv;jac69} on the datasets, which is a time consuming and difficult task not least because the number and type of chemical phases at each temperature is not known \textit{a priori}.  \fig{nmf_reaction_pathway_tracking} shows the results obtained by \nmfa (a), compared to those obtained by carrying out a multi-phase Rietveld refinement on each of the 25 datasets (b).

We found we got the best results when we ran \nmfa with 4~components.
However, the chemistry suggested that there may be 6~chemical species.
By carrying out Rietveld refinements on the NMF components we could attribute mixtures of multiple chemical species to each component.  The plot in \fig{nmf_reaction_pathway_tracking}(a) was then obtained by multiplying the NMF component weights by the chemical weights in each NMF component to obtain the evolution of chemical species with temperature.  The agreement with the chemical mol percentages obtained by carrying out laborious multi-phase Rietveld refinements on each of the datasets is very good. This shows that the NMF decomposition is yielding chemically sensible components despite it not using any chemical information to carry out the decomposition.

\section{Conclusion}

    We report here the \nmfa app, which makes NMF more easily accessible to diffraction experimentalists, both at the beamline and at their home institution. This tool provides users with a fast, reliable, intuitive way to gain physically meaningful information about large datasets rapidly.  To use the app, simply create and log in to your account on the PDFitc platform at https://pdfitc.org and upload your set of data.

    End-users should remember that NMF is a purely mathematical construct. It is important to exercise good sense and physical intuition when interpreting scientific results from an  NMF.  However, it is emerging as a powerful approach to rapidly assess and extract insight from large sets of related PDF and diffraction data and we hope that \nmfa can be a useful tool to help in its application.


\ack{Acknowledgements}
  
We would like to thank Alexander Schöekel for assistance in use of beamline P02.1.

\ack{Funding information}

This work was supported as part of GENESIS: A Next Generation Synthesis Center, an Energy Frontier Research Center funded by the U.S. Department of Energy, Office of Science, Basic Energy Sciences under Award Number DE-SC0019212. Sample preparation and PDF data analysis in the Ravnsbaek Group was funded by the Carlsberg Foundation (Grant CF17-0823) and data acquisition at DESY by Danscatt.

     Data for the chemical synthesis example was acquired at the 28-ID-2 beamline of the National Synchrotron Light Source II, a US DOE Office of Science User Facility operated for the DOE Office of Science by Brookhaven National Laboratory under contract No. DE-SC0012704. Data for the \textit{operando} battery experiment were collected at beamline P02.1, PETRA III, DESY (Hamburg, Germany), which is member of the Helmholtz Association HGF.

\bibliographystyle{iucr}
\bibliography{zt_nmf_itc}

\end{document}